\documentclass{article}
\usepackage{graphicx}
\usepackage{amsmath}
\usepackage{amsfonts}
\usepackage{amssymb}
\begin{document}

\title{Two remarks on wetting and emulsions}
\author{P.\ G.\ de Gennes\\Coll\`{e}ge de France, 11 place M.\ Berthelot,\\75231 Paris Cedex, France}
\maketitle
\begin{abstract}
This paper is extracted from an opening address given at the workshop
''Wetting: from microscopic origins to industrial applications'' (Giens, May
6-12, 2000). It discusses two special points \ \ a) the nature of line
energies for a contact line \ \ b) the aging of emulsions.
\end{abstract}

\section{Introduction}

These notes are dedicated to the memory of Karol Mysels.\ I met him only
rather late (in the 1980's).\ But I learned a large number of important things
from him.\ I also treasure a certain group of his original slides on soap
films, which he gave to me, and which I have used in a number ($\sim200)$ of
talks in high schools.\ Karol was both a gifted experimentalist and a deep
thinker.\ It was one of my great prides to have Karol and Estella as
participants for the Stockholm ceremonies, in which I was involved in 1991.

The following notes are a (clumsy) attempt to contribute to some things in
which Karol was deeply interested: the basics of wetting, and the stability of
foams and emulsions.\ The notes are rough, and would have been much improved
if he had still been with us...\ We shall not forget him.

\section{Line energies}

If we pinch a violin string (fig.\ 1), we obtain a\ static triangular
form.\ This is associated with a standard line tension $\Im$. For small
displacements $u$ with Fourier\ transforms $u_{q}$ the energy has the standard form:%

\begin{equation}
E=\underset{q}{\Sigma}\frac{1}{2}\Im q^{2}\left|  u_{q}\right|  ^{2}%
\end{equation}%

%TCIMACRO{\FRAME{fhF}{1.6829in}{0.8034in}{0pt}{}{}{00_0601_fig1.eps}%
%maintain-aspect-ratio TRUE;  display "PICT";  valid_file "F";
%width 1.6829in;  height 0.8034in;  depth 0pt;  original-width 4.1407in;
%original-height 1.644in;  cropleft "-0.0024109";  croptop "1.096791";
%cropright "0.9973891";  cropbottom "-0.082409";
%filename '00_0601_fig1.eps';file-properties "XNPEU";}}}%
%BeginExpansion
\begin{figure}
[h]
\begin{center}
\includegraphics[
trim=-0.009983in -0.135480in 0.010811in -0.159124in,
height=0.8034in,
width=1.6829in
]%
{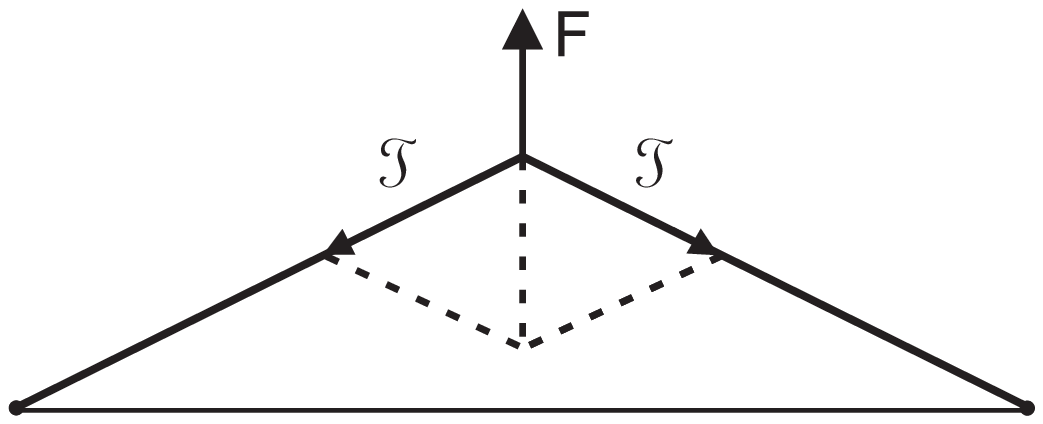}%
\end{center}
\end{figure}
%EndExpansion

On the other hand, if we pinch a contact line (for instance between a fluid, a
solid, and air) we find a very different form (fig. 2). This originates from a
different structure of the energy:%

\begin{equation}
E=\underset{q}{\Sigma}\frac{1}{2}\gamma f(\theta_{e})\left|  q\right|  \left|
u_{q}\right|  ^{2}%
\end{equation}

where $\theta_{e}$ is the equilibrium contact angle, $\gamma$ is the surface
tension, and $f(\theta_{e})$ is a dimensionless function, discussed in refs
$\cite{joanny}$ and \cite{brochard}.\ The $\left|  q\right|  $ dependance is
very singular: it comes from the integration of $q^{2}$ contributions over the
width of the perturbed fluid region ($q^{-1}$).%

%TCIMACRO{\FRAME{fhF}{1.8836in}{1.4486in}{0pt}{}{}{00_0601_fig2.eps}%
%maintain-aspect-ratio TRUE;  display "PICT";  valid_file "F";
%width 1.8836in;  height 1.4486in;  depth 0pt;  original-width 6.186in;
%original-height 4.7366in;  cropleft "0";  croptop "1";  cropright "1";
%cropbottom "0";  filename '00_0601_fig2.eps';file-properties "XNPEU";}}}%
%BeginExpansion
\begin{figure}
[h]
\begin{center}
\includegraphics[
height=1.4486in,
width=1.8836in
]%
{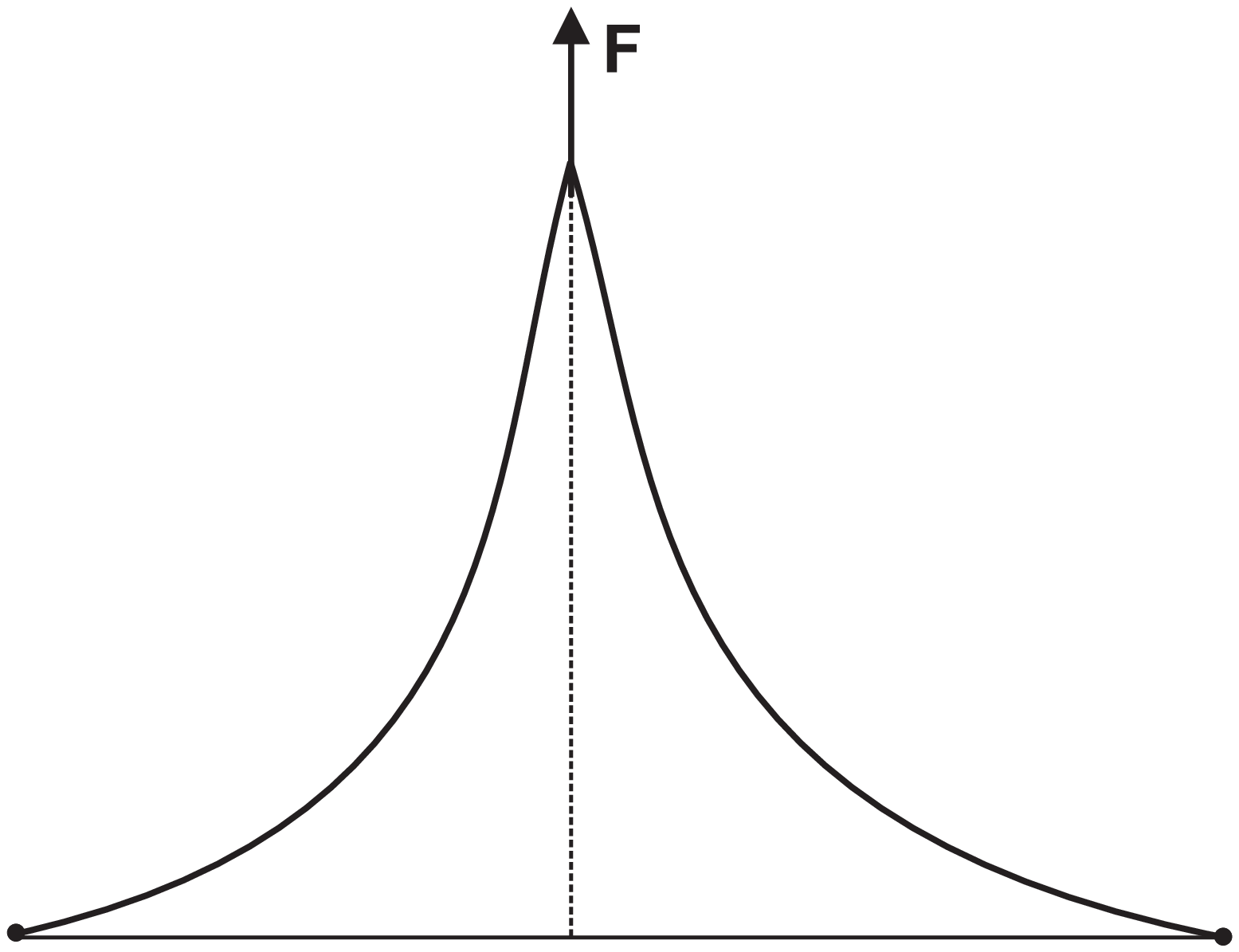}%
\end{center}
\end{figure}
%EndExpansion

Eq.\ 2 corresponds to what I call ''fringe elasticity''.\ It holds for scales
$q^{-1}$ which are smaller than the Laplace length:%

\begin{equation}
\varkappa^{-1}=\left(  \frac{\gamma}{\rho g}\right)  ^{1/2}%
\end{equation}

$\rho$: density, $g$: gravitational acceleration.

It is well known that we should add in equation (2) an intrinsic line tension
$\Im_{0}$, transforming eq. (2) into:%

\begin{equation}
E=\underset{q}{\Sigma}\left\{  \frac{1}{2}\gamma f(\theta_{e})\left|
q\right|  +\Im_{0}q^{2}\right\}
\end{equation}

Molecular models give values of $\Im_{0}$ (positive or negative) which are of order:%

\begin{equation}
\Im_{0}=\gamma a
\end{equation}

where $a$ is a molecular size (a few angstroms).

At submillimeter scales ($a<q^{-1}<\varkappa^{-1}$) this correction should be
unobservable.\ However, a number of groups \cite{neuman} \cite{drelich 96}
\cite{drelich 97} \cite{amirfazli} \cite{wang} have tried to measure $\Im_{0}$
by looking at the contact angle in small droplets.\ The result can be stated
in the form:%

\begin{equation}
\Im_{0}=\gamma\ell
\end{equation}

where $\left|  \ell\right|  $ is often anomalously large (of order 1 micron).

My own view is that this is an artefact of optical methods.\ One example,
based on ray optics, is illustrated on fig.\ 3. In principle, we determine
$\theta$ through the maximum deflection angle of a reflected ray
$\varphi=2\theta_{e}$.\ However, because of diffraction effects, the actual
last measurement point is at a distance $\sim\lambda$ from the contact line,
and has a weaker deflection $\varphi-\epsilon$, where $\epsilon\sim\lambda/R$
($R$: radius of curvature; $\lambda$: optical wavelength).\ This correction
has exactly the same structure as the inclusion of $\Im_{0}$, and gives an
apparent $\Im_{0}:$%

\begin{equation}
\Im_{0}/app\sim\gamma\lambda
\end{equation}%

%TCIMACRO{\FRAME{fhF}{2.079in}{1.081in}{0pt}{}{}{00_0601_fig3.eps}%
%maintain-aspect-ratio TRUE;  display "PICT";  valid_file "F";  width 2.079in;
%height 1.081in;  depth 0pt;  original-width 5.1275in;
%original-height 2.6333in;  cropleft "0";  croptop "1";  cropright "1";
%cropbottom "0";  filename '00_0601_fig3.eps';file-properties "XNPEU";}}}%
%BeginExpansion
\begin{figure}
[h]
\begin{center}
\includegraphics[
height=1.081in,
width=2.079in
]%
{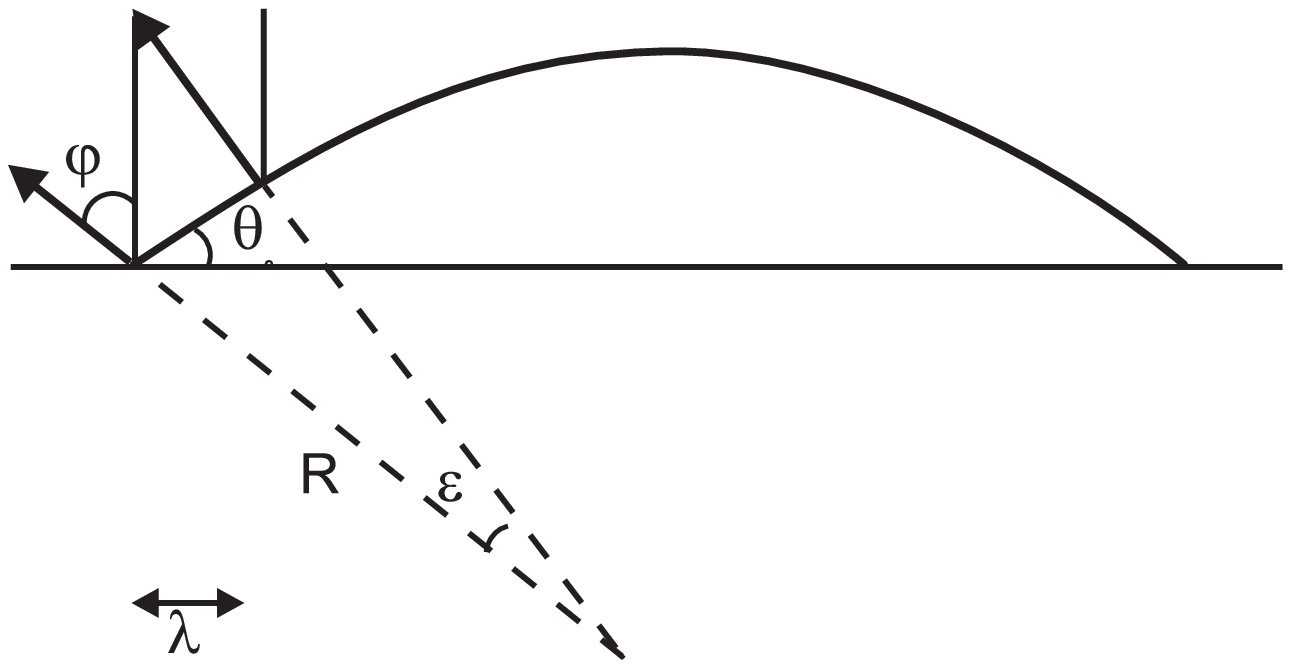}%
\end{center}
\end{figure}
%EndExpansion

This is the artefact.\ I believe that all optical methods have the same defect.

\section{Aging of emulsions}

Certain $O/W$ emulsions, which are monodisperse, can coalesce slowly, with
droplets growing in size, but remaining monodisperse $\cite{deminière}.\;$This
occurs in the absence of contaminants.\ In practice, many practical emulsions
do contain some small particles, with a wettability such that the particle
gets hooked at the $O/W$ interface, although it is rather hydrophobic. These
grains induce the rupture of oil films by the so called Garrett process
$\cite{garrett}.\;$If the grains are very dilute (so that each grain works
individually), this can lead to a rough, polydisperse emulsion.\ Our aim here,
is to discuss this, at the level of scaling laws, following the lines of ref
.$\cite{degennes}.$

Start with a single grain: when trapped on a film, it induces rupture within a
certain time $\tau$.\ When the film is locally destroyed, we postulate that
the particle \ binds to a neighboring interface -the size of the jump being
comparable to the diameter of the initial droplets.

We are thus led to think first of a random walk with a diffusion coefficient:%

\begin{equation}
D=d^{2}/\tau
\end{equation}

This walk destroys some droplets: the overall amount of interface present
decreases, and this could alter the control parameters: we want to avoid this
to reach a simple discussion:

\qquad a) We assume that there is a plentiful reservoir of surfactant
(concentration above $cmc$).

\qquad b) The thickness of the water films could increase, and this would
reduce the coalescence rates.\ If we estimate the diffusion coefficient
$D_{W}$ of \textit{water} in the structure, we find that it is related to
Poiseuille films in the Plateau borders (of diameter $h$) induced by gradients
of the pressure:%

\begin{equation}
p\cong p_{0}-\gamma/h
\end{equation}

The (qualitative) result is:%

\begin{equation}
D_{W}\cong\frac{\gamma}{\eta_{W}}h
\end{equation}

where $\eta$ is the water viscosity. In all what follows, we assume
$D_{W}>>D.$ Then water can diffuse fast from the ''wounded'' region to the
bulk of the monodisperse emulsion. It is then plausible to assume that the
water films remain at constant thickness.

A major point is the following: the grain does \textit{not} follow an
arbitrary random walk in the emulsion.\ If it ''digs a tunnel'' in the network
of droplets, the Laplace pressure at the bottom of the tunnel forces it to
retract very soon.

This retraction would follow the Washburn law for fluid motion in a capillary,
with a certain diffusion coefficient $D_{r}$ ($r$ stands for retraction).\ We
assume (as in agreement with usual conditions) that $D_{r}>D_{W}>D$.

In this situation, the grain can move only at the interface between a coarse
oil drop and the unperturbed fine emulsion, as shown on fig.\ 4.\newpage%

%TCIMACRO{\FRAME{fhF}{1.4019in}{1.4866in}{0pt}{}{}{00_0601_fig4.eps}%
%maintain-aspect-ratio TRUE;  display "PICT";  valid_file "F";
%width 1.4019in;  height 1.4866in;  depth 0pt;  original-width 4.5818in;
%original-height 4.862in;  cropleft "0";  croptop "1";  cropright "1";
%cropbottom "0";  filename '00_0601_fig4.eps';file-properties "XNPEU";}}}%
%BeginExpansion
\begin{figure}
[h]
\begin{center}
\includegraphics[
height=1.4866in,
width=1.4019in
]%
{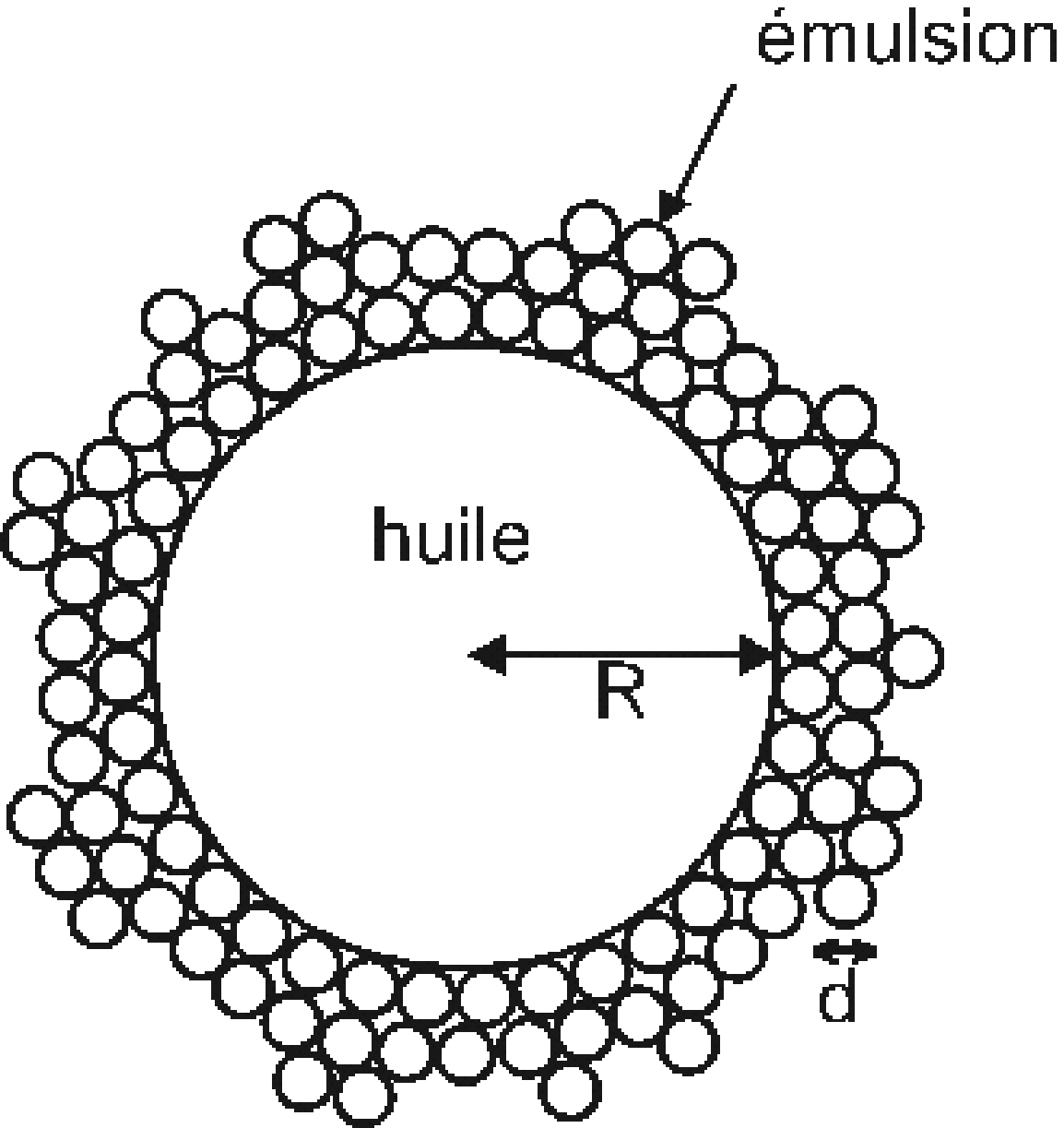}%
\end{center}
\end{figure}
%EndExpansion

It builds up a coarse drop of radius $R(t)$ increasing with time $t$. The
number of ruptured droplets is $t/\tau(>>1)$, and the oil volume of the coarse
drop is (omitting coefficients):%

\begin{equation}
R^{3}=\frac{t}{\tau}d^{3}=dDt
\end{equation}

During time $t$, the grain has moved at random on the surface of the coarse
drop, spanning a length $s$ such that $s^{2}=Dt$.\ We see from this that:%

\begin{equation}
\frac{\ell^{2}}{R^{2}}=\frac{R}{d}>>1
\end{equation}

This means that the particle explores fast the surface, and ensures that the
coarse droplet is essentially spherical.\ (For more details on the roughness
of the interface, see ref.\cite{degennes}$.$

Our conclusions are the following:

If we start, at $t=0,$ with a monodisperse emulsion, containing a very dilute
suspension of active grains, we expect a first stage, where each grain
generates one coarse droplet of size $R(t)$, growing like $t^{1/3}.$

At a certain moment ($t=t^{\ast}),$ these coarse droplets enter into
contact.\ At $t>t^{\ast},$ the original fine emulsion is not the main
component, and the growth laws must become more complex. Even if $D_{W}$ is
fast, the water films will thicken, and coalescences should slow down, or even
stop, as often observed.\ On the other hand, if $\tau$ has become long, one
could possibly arrive at a fusion of the coarse droplets, leading to an
inverted ($W/O$) emulsion.

On the whole, we are still very far from understanding the aging of
emulsions.\ But we clearly have an intrinsic blow up of the walls (which here
are water films) plus an extrinsic process due to grains (or other external
objects). Monodisperse emulsions may allow us to separate neatly the two processes.\newpage

\ \ \ \

\end{document}